\documentclass[altaffilletter,aps,nofootinbib,twocolumn,prd,eqsecnum,preprintnumbers,superscriptaddress]{revtex4-1}
\pdfoutput=1
\usepackage[caption=false]{subfig}
\usepackage{graphicx}
\usepackage{amsmath}
\usepackage{amsfonts}
\usepackage{amssymb}
\usepackage{color}
\usepackage{bm}
\usepackage{mathrsfs}
\usepackage{epstopdf}
\usepackage{url}
\usepackage{footnote}
\usepackage{textcomp}
\usepackage{dsfont}
\usepackage{ulem}
\usepackage{hyperref}
\usepackage{enumerate}

\makeatletter
\newcommand*{\rom}[1]{\expandafter\@slowromancap\romannumeral #1@}

\makeatother

\begin{document}

\title{Black Hole Perturbations and Quasinormal Modes in Hybrid Metric-Palatini Gravity}

\author{Che-Yu Chen}
\email{b97202056@gmail.com}
\affiliation{Institute of Physics, Academia Sinica, Taipei 11529, Taiwan}
\affiliation{Department of Physics and Center for Theoretical Sciences, National Taiwan University, Taipei, Taiwan 10617}
\affiliation{LeCosPA, National Taiwan University, Taipei, Taiwan 10617}

\author{Yu-Hsien Kung}
\email{r06222010@g.ntu.edu.tw}
\affiliation{Department of Physics and Center for Theoretical Sciences, National Taiwan University, Taipei, Taiwan 10617}
\affiliation{LeCosPA, National Taiwan University, Taipei, Taiwan 10617}

\author{Pisin Chen}
\email{pisinchen@phys.ntu.edu.tw}
\affiliation{Department of Physics and Center for Theoretical Sciences, National Taiwan University, Taipei, Taiwan 10617}
\affiliation{LeCosPA, National Taiwan University, Taipei, Taiwan 10617}
\affiliation{Kavli Institute for Particle Astrophysics and Cosmology, SLAC National Accelerator Laboratory, Stanford University, Stanford, CA 94305, USA}
\begin{abstract}
The rapid advancement of gravitational wave astronomy in recent years has paved the way for the burgeoning development of black hole spectroscopy, which enhances the possibility of testing black holes by their quasinormal modes (QNMs). In this paper, the axial gravitational perturbations and the QNM frequencies of black holes in the hybrid metric-Palatini gravity (HMPG) are investigated. The HMPG theory is characterized by a dynamical scalar degree of freedom and is able to explain the late-time accelerating expansion of the universe without introducing any \textit{ad hoc} screening mechanism to preserve the dynamics at the Solar System scale. We obtain the master equation governing the axial gravitational perturbations of the HMPG black holes and calculate the QNM frequencies.  Moreover, in the scrutiny of the black holes and their QNMs, we take into account the constraints on the model parameters based on the post-Newtonian analysis, and show how the QNM frequencies of the HMPG black holes would be altered in the observationally consistent range of parameter space.
\end{abstract}

\maketitle

\section{Introduction}

The recent direct detection of gravitational waves \cite{Abbott:2016blz,LIGOScientific:2018mvr} is undoubtedly a major breakthrough in the field of modern physics, because it not only verifies the existence of gravitational waves predicted by Einstein's general relativity (GR), but also ushers in a new era of gravitational wave astronomy. In particular, the gravitational waves emitted from black hole mergers typically contain a wealth of information regarding the nature of spacetime under strong gravitational fields, which is usually unattainable with electromagnetic observations only. Furthermore, one can also make use of these powerful gravitational wave \textit{telescopes} to examine whether our current understanding of black hole physics based on GR should be revised, hence to test the underlying theories of gravity.  

A typical merger event of a binary black hole system consists of three stages \cite{Abadie:2011kd,Berti:2015itd}. The first one is the inspiral stage, during which the two black holes rotate around each other, and the properties in this process, including the emitted gravitational waves (known as the \textit{chirp} signals), can be well-approximated by post-Newtonian methods. The second stage is the merger stage. At this stage, the gravitational field is extremely strong and the process can only be modeled numerically. The last stage is the ringdown stage during which the final black hole formed after the merger gradually settles. The distortion of the final black hole in shape undergoes decaying oscillations, which are essentially a superposition of several modes, called quasinormal modes (QNMs). Due to the dissipative nature of the system at this stage, the frequencies of QNMs are complex-valued, whose real part describes the oscillation, and the imaginary part corresponds to the exponential decay of the amplitudes. The ringdown stage can be essentially described using the theory of black hole perturbations. Remarkably, the QNM frequencies only depend on the parameters that describe the final black holes and they are independent of how the modes are triggered in the first place. Therefore, black hole QNMs can be a very powerful tool to test the black hole no-hair theorem as well as the underlying theories of gravity. This is the focus of this paper. We would like to refer the readers to Refs.~\cite{Kokkotas:1999bd,Berti:2009kk,Konoplya:2011qq} for reviews about black hole QNMs.

Among the plethora of gravitational theories, we will consider the black hole perturbations and the QNMs in the hybrid metric-Palatini gravity (HMPG) \cite{Harko:2011nh}, in which the gravitational action consists of the standard Einstein-Hilbert term constructed solely by the metric $g_{\mu\nu}$, as well as a function of another Ricci scalar which is defined based on the Palatini variation principle. One of the important features of HMPG is that the theory allows the existence of a long-range dynamical scalar field, such that the scalar field can drive the accelerating expansion of the universe at the cosmological scale. Also, as will be shown later, it is possible that the scalar field does not alter the dynamics at the local Solar System scale. Therefore, it seems not necessary to introduce any additional screening mechanism in this theory \cite{VargasdosSantos:2017ggl}, which is usually needed in other theories such as the metric-$f(R)$ gravity \cite{Sotiriou:2008rp,DeFelice:2010aj}. In addition, unlike the Palatini-$f(R)$ gravity \cite{Olmo:2011uz}, the HMPG theory contains one dynamical scalar degree of freedom{\footnote{In the literature, the HMPG theory is also called $f(X)$ gravity, where $X$ is a dynamical scalar degree of freedom quantifying the non-zeroness of the trace of the Einstein equation \cite{Capozziello:2012ny}.}}, and the theory does not suffer from the microscopic instabilities in the Palatini-$f(R)$ gravity.

Ever since it was proposed, the HMPG theory has received wide attention. In cosmology, some interesting cosmological solutions have been found \cite{Capozziello:2012ny,Boehmer:2013oxa}, inflation \cite{Kausar:2019iwu} and cosmological perturbations \cite{Lima:2014aza} have been studied, and the dynamical analysis of HMPG cosmology has been carried out \cite{Carloni:2015bua}. The HMPG theory can be tested from cosmological observations \cite{Lima:2015nma,Lima:2016npg,Leanizbarrutia:2017xyd}. In astrophysics, it has been shown that the HMPG scalar field could mimic the behavior of dark matter \cite{Capozziello:2012qt,Capozziello:2013uya}, especially describe the flat region in galactic rotation curves \cite{Capozziello:2013yha}. In addition, the HMPG theory is able to support wormhole geometries \cite{Capozziello:2012hr,Lobo:2012ai,Korolev:2020ohi}. Very recently, the black hole solutions \cite{Danila:2018xya,Bronnikov:2019ugl,Bronnikov:2020vgg}, string-like objects \cite{Harko:2020oxq,Bronnikov:2020zob}, and other compact stellar objects \cite{Danila:2016lqx} in HMPG have also been investigated. The Cauchy problem \cite{Capozziello:2013gza}, the Noether symmetry \cite{Borowiec:2014wva}, the post-Newtonian analysis \cite{Dyadina:2018jrk,Dyadina:2019dsu}, and constraints from stellar motions \cite{Borka:2015vqa} have been studied. Furthermore, the HMPG theory can be extended by including higher dimensions \cite{Fu:2016szo} and torsion fields \cite{Capozziello:2013dja}, with the motivation of the latter being to consider spinors. Recently, the hybrid formalism has been shown to preserve the Weyl symmetry when trying to unify the theory with standard model particles \cite{Edery:2019txq}.  Also, a natural extension of the HMPG theory is to consider a general function of the two Ricci scalars \cite{Flanagan:2003iw,Tamanini:2013ltp}. This generalized HMPG theory contains two additional scalar degrees of freedom and has been explored in cosmology \cite{Rosa:2017jld,Bombacigno:2019did,Rosa:2019ejh,Sa:2020qfd} and astrophysics \cite{Rosa:2018jwp,Rosa:2020uoi}, although its hybrid structure may cause unwanted instabilities \cite{Koivisto:2013kwa,Jimenez:2020dpn}. For the review on HMPG, we refer the readers to Refs.~\cite{Capozziello:2015lza,Harko:2020ibn}.

In this paper, we will adopt the scalar-tensor representation of the HMPG theory. Due to the complexity of the field equations, we will follow the work in Ref.~\cite{Danila:2018xya} to obtain the black hole solutions using numerical integrations. As opposed to the Palatini-$f(R)$ gravity, whose QNMs of charged black hole perturbations were studied in Refs.~\cite{Chen:2018mkf,Chen:2018vuw}, the HMPG theory contains a dynamical scalar degree of freedom and it is possible to obtain vacuum spacetimes different from their GR counterparts. We will focus on the spherically symmetric black hole spacetime in which both Ricci scalars are zero but the Ricci tensor is generically not. The black hole solutions will be obtained under the consideration of the requirements from the post-Newtonian analysis. We will focus on the axial gravitational perturbations of the HMPG black holes and derive the master equation. The QNM frequencies will be evaluated using the 6th order Wentzel-Kramers-Brillouin (WKB) method \cite{Konoplya:2019hlu}.

This paper is outlined as follows. In Sec.~\ref{sec.HMPG}, we briefly review the formulation of the HMPG theory and its scalar-tensor representation. The previously obtained results of the post-Newtonian analysis are reviewed as well. In Sec.~\ref{sec.BH}, the HMPG black holes are studied numerically. In Sec.~\ref{sec.QNM}, we derive the master equation governing the axial gravitational perturbations for the HMPG black holes and calculate the QNM frequencies. We finally conclude in Sec.~\ref{sec.conclu}.

\section{HMPG formulation}\label{sec.HMPG}
In this section, we will briefly review the HMPG theory \cite{Harko:2011nh}, including its action, equations of motion, and some of its important properties. We will demonstrate that the theory can be recast into the scalar-tensor representation. It turns out that the calculations that we are going to go through in this paper are more straightforward within this representation. Then, we will review the post-Newtonian analysis of HMPG and present its main results \cite{Harko:2011nh,Dyadina:2018jrk,Dyadina:2019dsu}. The results of the post-Newtonian analysis not only exhibit the motivations of considering HMPG, but also turn out to be important in our later usage when investigating black hole perturbations in the theory.

\subsection{HMPG and its scalar-tensor representation}
The action of HMPG is written as \cite{Harko:2011nh}
\begin{align}
    S = \frac{1}{2\kappa^2} \int d^4 x \sqrt{-g} \left[ R + f(\mathcal{R})\right] + S_m\,,\label{action}
\end{align}
where $S_m \equiv \int d^4x \sqrt{-g}\mathcal{L}_m$ is the matter action and $\kappa \equiv 8\pi G$. The Ricci scalar $R$ is constructed solely by the metric $g_{\mu\nu}$ and it stands for the Einstein-Hilbert term in the gravitational action. The second term in the action is a function of another curvature invariant $\mathcal{R} \equiv g^{\mu\nu}\mathcal{R}_{\mu\nu}$, which comes from the contraction between the metric and the Ricci tensor constructed solely by the independent affine connection $\hat\Gamma$, namely 
\begin{align}
    \mathcal{R}_{\mu\nu} \equiv \partial_\sigma \hat \Gamma^{\sigma}_{\mu\nu} - \partial_\nu \hat \Gamma^{\sigma}_{\mu\sigma} + \hat \Gamma^{\sigma}_{\sigma\rho} \hat \Gamma^{\rho}_{\mu\nu} - \hat \Gamma^{\sigma}_{\mu\rho} \hat \Gamma^{\rho}_{\sigma\nu}\,.
\end{align}
Therefore, the gravitational modifications on top of the Einstein-Hilbert action are contributed by the addition of $f(\mathcal{R})$ term.

Since the action contains an affine connection which is independent of the metric $g_{\mu\nu}$, one has to vary the action with respect to them separately to derive the equations of motion. After varying the action \eqref{action} with respect to the metric, we can directly obtain the field equation
\begin{align}
    G_{\mu\nu} + f_\mathcal{R}\mathcal{R}_{\mu\nu} - \frac{1}{2}f(\mathcal{R})g_{\mu\nu} = \kappa^2 T_{\mu\nu} \,,
\end{align}
where $G_{\mu\nu}$ is the Einstein tensor defined by $g_{\mu\nu}$, and $f_\mathcal{R} \equiv d f(\mathcal{R})/ d \mathcal{R}$. The energy-momentum tensor $T_{\mu\nu}$ is defined as
\begin{align}
    T_{\mu\nu} \equiv - \frac{2}{\sqrt{-g}} \frac{\delta \left(\sqrt{-g}\mathcal{L}_m \right)}{\delta g^{\mu\nu}}\,,
\end{align}
where the matter is assumed to couple with the metric $g_{\mu\nu}$ only. Besides, after varying the action with respect to the independent connection, we get
\begin{align}
    \hat \nabla_\sigma \left( \sqrt{-g} f_\mathcal{R} g^{\mu\nu} \right) = 0 \,.
    \label{compatible}
\end{align}
In the above expression, $\hat\nabla$ denotes the covariant derivative defined by the affine connection $\hat\Gamma$. Eq.~\eqref{compatible} implies that the independent connection is compatible with an auxiliary metric $q_{\mu\nu}\equiv f_\mathcal{R} g_{\mu\nu}$, which is conformal to $g_{\mu\nu}$. Therefore, the HMPG theory looks like a bi-metric theory, with a physical metric $g_{\mu\nu}$ and an auxiliary metric $q_{\mu\nu}$, while only one scalar degree of freedom is involved due to the conformal relation between the metrics. Note that in the metric-$f(R)$ and the Palatini-$f(R)$ theories, one can also define an auxiliary metric conformal to the physical metric. However, due to the hybrid structure of the HMPG theory, the HMPG theory is completely different from the other two kinds of $f(R)$ theories, and it acquires distinctive physical properties, as will be shown later.

Similar to the metric-$f(R)$ and Palatini-$f(R)$ gravity, the HMPG theory can also be formulated in a scalar-tensor representation \cite{Harko:2011nh}. By introducing an auxiliary field $\chi$, the action \eqref{action} can be rewritten as
\begin{align}
    S = \frac{1}{2\kappa^2} \int d^4x \sqrt{-g} \left[ R + f\left(\chi\right)+ f_\chi\left(\mathcal{R} - \chi\right) \right] + S_m \,,
    \label{actionscalartensor}
\end{align}
where $f_\chi\equiv df(\chi)/d\chi$. By varying the action \eqref{actionscalartensor} with respect to $\chi$, we find that $\chi = \mathcal{R}$. Thus, if $d^2f/d\mathcal{R}^2 \neq 0$, the field $\chi$ is dynamically equivalent to the Ricci scalar $\mathcal{R}$. Furthermore, with the following definitions
\begin{align}
    \phi \equiv f_\chi\,, \quad V(\phi) \equiv \chi f_\chi - f(\chi) \,,
\end{align}
we obtain the scalar-tensor representation of the HMPG theory
\begin{align}
    S = \frac{1}{2\kappa^2} \int d^4x \sqrt{-g} \left[ R + \phi \mathcal{R} - V(\phi) \right] + S_m \,.
    \label{scalaraction}
\end{align}
The equations of motion of HMPG in the scalar-tensor representation can be obtained by varying the action \eqref{scalaraction} with respect to the metric $g_{\mu\nu}$, the scalar field $\phi$, and the affine connection: 
\begin{align}
    R_{\mu\nu} + \phi \mathcal{R}_{\mu\nu} - \frac{1}{2} \left[ R + \phi \mathcal{R} - V(\phi) \right] g_{\mu\nu}  &= \kappa^2 T_{\mu\nu}\,,
    \label{fieldequation}\\
    \mathcal{R} - V_{\phi} &= 0\,,
    \label{fieldequation2}\\
    \hat \nabla_{\sigma} \left( \sqrt{-g} \phi g^{\mu\nu} \right) &= 0 \,,
    \label{compatible2}
\end{align}
respectively. Since Eqs.~\eqref{compatible} and \eqref{compatible2} imply that the affine connection is the Levi-Civita connection of the auxiliary metric $q_{\mu\nu}\equiv\phi g_{\mu\nu}$, which is conformally related to the original metric $g_{\mu\nu}$, one can obtain the relation between $R_{\mu\nu}$ and $\mathcal{R}_{\mu\nu}$ as follows
\begin{align}
    \mathcal{R}_{\mu\nu} = R_{\mu\nu} + \frac{3}{2\phi^2}\partial_\mu \phi \partial_\nu \phi - \frac{1}{\phi} \left(\nabla_\mu \nabla_\nu \phi + \frac{1}{2}g_{\mu\nu} \Box \phi  \right) \,.
    \label{RRrelation}
\end{align}
Note that the covariant derivative $\nabla_\mu$ here is constructed from the metric $g_{\mu\nu}$. The two Ricci scalars are thus related by
\begin{align}
    \mathcal{R} = R + \frac{3}{2\phi^2} \partial_\mu \phi \partial^\mu \phi - \frac{3}{\phi} \Box \phi\,. \label{RRrelation2}
\end{align}
Replacing the Ricci scalar $\mathcal{R}$ in Eq.~\eqref{scalaraction} with the relation \eqref{RRrelation2}, the action \eqref{scalaraction} can be rewritten as
\begin{align}
    S =&\, \frac{1}{2\kappa^2}\int d^4x \sqrt{-g} \left[ \left( 1+ \phi \right) R + \frac{3}{2\phi} \partial_\mu \phi \partial^\mu \phi - V(\phi) \right] \nonumber\\&+ S_m\,.\label{finalactionst}
\end{align}
It should be emphasized that the action \eqref{finalactionst} is very similar to that of the Palatini-$f(R)$ gravity in its Brans-Dicke representation. The only difference is that in the Palatini-$f(R)$ gravity, the coupling between the scalar field and the Ricci scalar $R$ appears in the form of $\phi R$, while in the action \eqref{finalactionst} it appears as $(1+\phi)R$.{\footnote{Note that the theory described by the action \eqref{finalactionst} actually belongs to the Bergmann-Wagoner-Nordtvedt types of scalar-tensor theories \cite{Bergmann:1968ve,Wagoner:1970vr,Nordtvedt:1970uv}.}} It will be shown later that this slight difference in the scalar-curvature coupling would render the HMPG theory a distinctive feature as compared with the Palatini-$f(R)$ gravity. 

In order to have a clearer picture of how the scalar field $\phi$ modifies GR and changes the Einstein equations, we follow the procedure in Ref.~\cite{Danila:2018xya} and rewrite Eq.~\eqref{fieldequation} by using the relation \eqref{RRrelation}. One can obtain the following equation
\begin{align}
    G_{\mu\nu} = \kappa^2T_{\mu\nu}^{\textrm{eff}}\,,\label{finalequationofmotionGT}
\end{align}
where the effective energy-momentum tensor is defined as
\begin{align}
T_{\mu\nu}^{\textrm{eff}} =& \frac{1}{1+\phi} \bigg\{ T_{\mu\nu} - \frac{1}{\kappa^2} \bigg[ \frac{1}{2} g_{\mu\nu}(V + 2 \Box \phi )- \nabla_\mu \nabla_\nu \phi  \nonumber\\
    &+\frac{3}{2\phi} \partial_\mu \phi \partial_\nu \phi - \frac{3}{4\phi} g_{\mu\nu}(\partial_\alpha \phi \partial^\alpha \phi) \bigg]\bigg\} \,.\label{eff}
\end{align}
Besides, by utilizing Eqs.~\eqref{fieldequation2} and \eqref{RRrelation2}, one can obtain the equation of motion of the scalar field as follows
\begin{align}
    - \Box \phi + \frac{1}{2\phi} \partial_\mu \phi \partial^\mu \phi + \frac{\phi[2V-(1+\phi)V_\phi]}{3} = \frac{\phi \kappa^2}{3} T \,,
    \label{effectkleingordon}
\end{align}
where $V_\phi\equiv dV/d\phi$ and $T\equiv g^{\mu\nu}T_{\mu\nu}$ stands for the trace of the energy-momentum tensor in the matter sector. 

An important feature indicated by the equations of motion \eqref{finalequationofmotionGT} and \eqref{effectkleingordon} is that unlike the Palatini-$f(R)$ theory, the scalar field $\phi$ in the HMPG theory is dynamical. In fact, in the Palatini-$f(R)$ gravity, the scalar field is non-dynamical and the field equations can be recast in an expression with non-trivial couplings between gravity and matter sectors. These couplings usually cause instabilities at microscopic scales \cite{Olmo:2011uz}. In contrast, the scalar field in the 
HMPG theory is dynamical and so it does not suffer from the microscopic instabilities mentioned above. Also, it can be proven that the energy-momentum tensor is conserved i.e., $\nabla_\mu T^{\mu\nu} = 0$, in the HMPG theory because the matter field only couples to the physical metric $g_{\mu\nu}$.

\subsection{The post-Newtonian analysis}
As we have mentioned in the Introduction, one of the advantages of the HMPG theory is that it can successfully describe the late-time expansion of our universe without altering the dynamics at the Solar System scale. This interesting feature can be directly appreciated with the post-Newtonian analysis. The post-Newtonian analysis of the HMPG theory was firstly presented together with the proposal of the theory itself \cite{Harko:2011nh}. Later, extensions to higher orders are carried out in Refs.~\cite{Dyadina:2018jrk,Dyadina:2019dsu}. Here, we shall briefly mention the results of the analysis.

To illustrate how the physics at the Solar System scale would be modified in the HMPG theory, one can study its associated post-Newtonian parameters \cite{Capozziello:2015lza}. In the weak-field and slow-motion limit, we consider a quasi-Minkowskian spacetime: $g_{\mu\nu}\approx\eta_{\mu\nu}+h_{\mu\nu}$ with $|h_{\mu\nu}|\ll1$ and $\phi=\phi_0+\phi_1(x)$, where $\phi_0$ is the value of the scalar field at the asymptotically distant region. The sub-leading order term of the scalar field $\phi_1(x)$ is assumed to be time independent due to the slow-velocity assumption. Expanding up to $\mathcal{O}(h^2)$, one can obtain the perturbed metric as follows \cite{Dyadina:2018jrk,Dyadina:2019dsu}:
\begin{align}
h_{00}&=\frac{2M}{\left(1+\phi_0\right)r}\left(1-\frac{\phi_0}{3}e^{-m_{\phi} r}\right)\,,\label{h00}\\
h_{ij}&=\frac{2M}{\left(1+\phi_0\right)r}\left(1+\frac{\phi_0}{3}e^{-m_{\phi} r}\right)\delta_{ij}\,,\label{hij}
\end{align}
where $M$ denotes the mass of the local object. On the above expression, the mass of the scalar field is defined via
\begin{equation}
m_{\phi}^2=\left[2V-V_\phi-\phi(1+\phi)V_{\phi\phi}\right]/3\big|_{\phi=\phi_0}\,.
\end{equation}
The perturbed scalar field $\phi_1$ can be written as
\begin{equation}
\phi_1=-\frac{\kappa^2}{12\pi}\frac{\phi_0M}{r}e^{-m_{\phi} r}\,.\label{varphi}
\end{equation}
Using the above equations, one can extract the effective gravitational constant and the post-Newtonian parameter $\gamma$ as
\begin{align}
    G_{\textit{eff}} &\equiv \frac{\kappa^2}{8\pi (1+\phi_0)} \left( 1 - \frac{\phi_0}{3} e^{- m_{\phi} r} \right)\,,\label{Geffconstant}\\
    \gamma &\equiv \frac{1 + (\phi_0/3) e^{- m_{\phi} r} }{1 - (\phi_0/3) e^{- m_{\phi} r}}\,.\label{gamma}
\end{align}

It can be seen from Eq.~\eqref{gamma} that for the HMPG theory to be consistent with the local gravitational tests at the Solar System scale, i.e., $\gamma\approx 1$, one may require a very massive scalar field $\phi$ as in the case of the metric-$f(R)$ theory. It is well-known that this assumption is not consistent with the requirement that the scalar field has to be long-rang in order to modify the cosmological dynamics. In the HMPG theory, it is possible to have a long-range scalar field while remain the dynamics in the local scale intact. This can be achieved by imposing a very small background field $\phi_0 \ll 1$ such that the magnitude of $m_{\phi}$, that is, the Yukawa-type correction, does not affect too much on $\gamma$. With a small asymptotic scalar field $\phi_0$, the theory is able to survive from local tests of gravity, and at the meantime the long-range Yukawa interaction induced by the scalar field can modify the cosmological dynamics of the universe.

At this point, we would like to emphasize that the results of the post-Newtonian analysis, especially Eqs.~\eqref{h00} and \eqref{hij} as well as the fact that $\phi_1$ has a prefactor $\phi_0$ (see Eq.~\eqref{varphi}), will be taken into account when we investigate the black hole solutions of HMPG and their perturbations later.


\section{Black hole solutions in HMPG}\label{sec.BH}

In order to investigate the black hole perturbations and QNMs in HMPG, one has to specify the black hole spacetime that is going to be perturbed. For the sake of simplicity, in this paper we will consider static and spherically symmetric black holes and investigate their gravitational perturbations. Also, in the rest of the paper, we will consider vacuum solutions ($T_{\mu\nu}=0$). It should be emphasized that in the presence of the dynamical scalar field, the static and spherically symmetric vacuum solution in HMPG could be different from the Schwarzschild solution. By focusing on the vacuum black hole solutions and their perturbations, we are able to directly understand how the black hole spacetime and QNMs are changed due to the geometrical corrections induced by the dynamical scalar field (or in other words, the corrections induced from the $f(\mathcal{R})$ modification in Eq.~\eqref{action}). Because of the complexity in the equations of motion, the black hole solutions in HMPG have been investigated purely numerically in Ref.~\cite{Danila:2018xya}, in which the authors focused on the spacetime outside the horizon and indicated the position of the event horizon by the existence of a Killing horizon for the timelike Killing vector in the metric tensor components{\footnote{It was pointed out very recently in Refs.~\cite{Bronnikov:2019ugl,Bronnikov:2020vgg} that in HMPG with a zero potential $V(\phi)$, asymptotically flat black holes with a single horizon do not exist generically. They may exist only in some special cases.}}. As will be shown later, the quasinormal modes of a black hole are determined by the spacetime property outside the horizon. Therefore, in this section we will adopt the procedures in Ref.~\cite{Danila:2018xya} and solve the black hole solutions numerically.

We will investigate the static and spherically symmetric solutions of the HMPG theory. The line element describing such a geometry can be generally represented by
\begin{align}
    ds^2 = -F(r) dt^2 + \frac{1}{G(r)}dr^2 + r^2 d\theta^2 +r^2\sin^2\theta d\varphi^2 \,.\label{SSSmetric}
\end{align}
The metric functions $F(r)$ and $G(r)$ only depend on the radial coordinate $r$. In the following calculations, we will focus on the spacetime outside the event horizon, namely, the range $r_{\mathcal{H}} < r < \infty$, where $r_{\mathcal{H}}$ is the event horizon of the black hole. Note that the metric functions $F(r)$ and $G(r)$ would vanish at the horizon.

Inserting the metric ansatz \eqref{SSSmetric} into the equations of motion \eqref{finalequationofmotionGT} and \eqref{effectkleingordon}, one obtains
\begin{align}
\frac{1}{r^2} \left( 1 - G -rG' \right) \left(1+\phi\right) &- G \left(\phi'' - \frac{3\phi'^2}{4\phi}\right) \nonumber\\
    - \frac{\phi'}{2r} \left(rG' + 4G\right) &-\frac{V(\phi)}{2} =0 \,,\label{tt}\\
\left[ \frac{1}{r^2}\left(G-1\right)+ \frac{GF'}{rF}\right]&\left(1+\phi\right) \nonumber\\
    + \phi'G \left(\frac{F'}{2F}+\frac{2}{r}+\frac{3\phi'}{4\phi}\right) &+ \frac{V(\phi)}{2} = 0 \,,\label{rr}
\end{align}
and
\begin{align}
     -G \left( \phi'' + \frac{F' \phi'}{2F} - \frac{\phi'^2}{2\phi} + \frac{2\phi'}{r} \right) &- \frac{1}{2}G' \phi' 
     \nonumber\\
     + \frac{\phi}{3} \left[ 2V-\left(1+\phi\right)V_{\phi}\right] &= 0 \,,\label{klein}
\end{align}
respectively, where the prime denotes the derivative with respect to $r$. Eqs.~\eqref{tt} and \eqref{rr} are the $tt$ and $rr$ components of the modified Einstein equation \eqref{finalequationofmotionGT}, respectively. Notice that the angular components of Eq.~\eqref{finalequationofmotionGT} are not provided here since they are proven to be redundant. For the sake of simplicity, we define a new function $u(r)\equiv\phi'(r)$. Then, one can obtain (see Ref.~\cite{Danila:2018xya} for more detail):
\begin{align}
    G'(r) =&\, \frac{1}{2r\phi (2+2\phi+ur)} \big[ 4\phi(1+\phi)(1-G)
    \nonumber\\
    &-2r^2\phi V - 8Gru\phi + 3Gr^2u^2 - 4Gr^2u'\phi \big]\,,\label{G'}\\
    F'(r) =&\, \frac{F}{2Gr\phi(2+2\phi+ur)} \big[ 4\phi(1+\phi)(1-G)
    \nonumber\\
    &- 2r^2\phi V - 8Gru\phi - 3Gr^2u^2 \big]\,,\label{F'}\\
    u'(r) =&\, \frac{u}{2} \left(\frac{u}{\phi} - \frac{F'}{F} - \frac{4}{r}\right) 
    \nonumber\\
    &+ \frac{1}{G} \left[ \frac{2\phi}{3} \left(V - \frac{1+\phi}{2}V_{\phi}\right) -\frac{G'u}{2} \right]\,.\label{u'}
\end{align}
After introducing a new variable $\xi = 1/r$, we have \cite{Danila:2018xya}
\begin{align}
    \frac{d\phi}{d\xi} &= -\frac{u}{\xi^2} \,,\label{diff1}\\
    -\xi^2 \frac{dG}{d\xi} &= \frac{\xi}{2\phi(2+2\phi+u/\xi)} \Bigg[ 4\phi(1+\phi)(1-G) 
    \nonumber\\
    &- \frac{2V\phi}{\xi^2} - \frac{8G\phi u}{\xi} + \frac{3Gu^2}{\xi^2} +4G\phi \frac{du}{d\xi} \Bigg] \,,\label{dG}\\
    -\xi^2 \frac{dF}{d\xi} &= \frac{F\xi}{2G\phi(2+2\phi+u/\xi)} \Bigg[ 4\phi(1+\phi)(1-G) 
    \nonumber\\
    &- \frac{2V\phi}{\xi^2} - \frac{8G\phi u}{\xi} - \frac{3Gu^2}{\xi^2} \Bigg]\,,\label{dF}\\
    -\xi^2 \frac{du}{d\xi} &= \frac{u}{2} \left(\frac{u}{\phi} + \frac{\xi^2}{F}\frac{dF}{d\xi} - 4\xi\right)
    \nonumber\\
    &+ \frac{1}{G} \left[ \frac{2\phi}{3} \left(V - \frac{1+\phi}{2}V_{\phi}\right) +\frac{\xi^2u}{2}\frac{dG}{d\xi} \right] \,.\label{du}
\end{align}
Furthermore, the derivatives of $F$ and $G$ appearing in Eq.~\eqref{du} can be substituted by using Eqs.~\eqref{dG} and \eqref{dF}. This yields
\begin{align}
    -\xi^2 G &\frac{du}{d\xi} = \frac{uG}{2} \left(\frac{u}{\phi} - 4\xi\right) -  \frac{u\xi}{2+2\phi+u/\xi} \times
    \nonumber\\
    &\left[ 2\left(1+\phi\right)\left(1-G\right) - \frac{V}{\xi^2} - \frac{4G u}{\xi}  +G\frac{du}{d\xi}\right] +\mathcal{V}(\phi) \,,\label{du2}
\end{align}
where
\begin{align}
    \mathcal{V}(\phi) \equiv \frac{2\phi}{3} \left(V - \frac{1+\phi}{2}V_{\phi}\right) \,.
\end{align}

In principle, after specifying the potential $V(\phi)$ and imposing the boundary conditions, we can use the set of differential equations \eqref{diff1}, \eqref{dG}, \eqref{dF}, and \eqref{du2} to solve the functions $F$, $G$, $\phi$, and $u$. At this point, we will further assume that the scalar field potential is zero, namely, $V(\phi)=0$. The spacetime is assumed to be asymptotically flat. For the black holes in the HMPG theory, the value of the scalar field potential at the asymptotic region $r\rightarrow\infty$ defines an effective cosmological constant. Therefore, the scalar field potential should approach zero when $r\rightarrow\infty$ for an asymptotically flat spacetime. A vanishing potential $V(\phi)=0$ turns out to be the simplest choice in the sense that the black hole, besides its mass, would then be completely described by the scalar field $\phi$ and its derivative $\phi'$ at the asymptotic region. If one includes a non-trivial potential, the parameters of the potential would come into play, hence enlarge the parameter space. In fact, when $V(\phi)=0$, it can be shown that the two Ricci scalars are identically zero, that is, $R=\mathcal{R}=0$. However, as has been shown in Ref.~\cite{Danila:2018xya,Bronnikov:2019ugl,Bronnikov:2020vgg}, the Ricci tensor would not vanish in general if the scalar field is dynamical. The black hole would differ from the Schwarzschild counterpart, even for $V(\phi)=0$.

The boundary conditions, on the other hand, should be imposed properly in order to respect the asymptotic flatness condition. More precisely, we have to assume the asymptotic value of the scalar field $\phi_0$, which is supposed to be very tiny according to the post-Newtonian constraints. Then, the metric functions $F$ and $G$ at the asymptotic region ($\xi\rightarrow0$), should also be imposed according to the post-Newtonian results, namely, Eqs.~\eqref{h00} and \eqref{hij}. Finally, the asymptotic value of the function $u$ can be determined by using Eq.~\eqref{varphi}. It should be noticed that at the asymptotic region where $\xi\rightarrow0$, the function $u$ can be approximated as $u\approx\phi_0\xi^2$, which is roughly proportional to $\phi_0$ and is extremely small{\footnote{We would like to mention that in the numerical calculations of Ref.~\cite{Danila:2018xya}, the asymptotic values of $\phi$ and $u$ are set independently. See FIGs~1 and 2 in Ref.~\cite{Danila:2018xya}.}}. After imposing these boundary conditions, the equations of motion describing the metric functions and the scalar field can be integrated numerically.

\section{Axial perturbations and QNMs}\label{sec.QNM}
As we have mentioned previously, the static and spherically symmetric vacuum solutions in HMPG are generically different from the Schwarzschild solution due to the presence of the dynamical scalar field. Even though we have assumed a zero potential $V(\phi)=0$ and the two Ricci scalars are identically zero, the dynamical scalar field would still alter the geometry and, in principle, leave some observational imprints with which one can distinguish them from their GR counterpart. In this section, we will focus on the axial gravitational perturbations of the aforementioned black hole solutions in HMPG and compute the corresponding QNM frequencies after we obtain the master equation governing the perturbations. The master equation will be derived by using the tetrad formalism. 

Without loss of generality, the perturbed metric of a static and spherically symmetric spacetime can be described by a non-stationary and axisymmetric metric whose symmetrical axis is tuned such that the spacetime metric does not depend on the azimuthal angle $\varphi$ \cite{Chandrabook}. In practice, an axisymmetric mode can be decomposed into a complete set of non-axisymmetric modes. At the level of linear approximations, the radial dependence of the modes is not affected by this angular decomposition \cite{Chandrabook}. Since the master equation of the modes is determined by the radial dependence in the mode decomposition, it is not affected by choosing a different polar axis in the coordinate system. This is essentially the same as in quantum mechanics why the radial wave function of an electron in a central field does not depend on the magnetic quantum number $m$.

If we consider only the axial perturbations, the perturbed spacetime metric $g_{\mu\nu}$ can be written as
\begin{align}
ds^2 = &-F(r) dt^2+r^2\sin^2\theta\left(d\varphi-\zeta dt-q_2dr-q_3d\theta\right)^2\nonumber\\
&+\frac{dr^2}{G(r)}+r^2d\theta^2\,.\label{perturbg}
\end{align}
On the above perturbed metric, the axial perturbations are encoded in the functions $\zeta$, $q_2$, and $q_3$. These functions, therefore, are functions of $t$, $r$, and $\theta$. Since we only focus on the axial gravitational perturbations which are basically some combinations of the functions $\zeta$, $q_2$, and $q_3$, the metric functions $F$ and $G$ are treated as the zeroth order quantities and they are functions of $r$ only. 

\subsection{Tetrad formalism}
We are going to derive the master equation by using the tetrad formalism \cite{Chandrabook}. The calculations within the tetrad formalism are based on the construction of a new tetrad frame, which is constructed by a basis $e^\mu_{(a)}$ associated with the original spacetime metric $g_{\mu\nu}$. Note that the tetrad indices are enclosed in parentheses to distinguish them from the tensor indices. The tetrad basis should satisfy
\begin{align}
e_{\mu}^{(a)}e^{\mu}_{(b)}&=\delta^{(a)}_{(b)}\,,\quad e_{\mu}^{(a)}e^{\nu}_{(a)}=\delta^{\nu}_{\mu}\,,\nonumber\\
e_{\mu}^{(a)}&=g_{\mu\nu}\eta^{(a)(b)}e^{\nu}_{(b)}\,,\nonumber\\
g_{\mu\nu}&=\eta_{(a)(b)}e_{\mu}^{(a)}e_{\nu}^{(b)}\equiv e_{(a)\mu}e_{\nu}^{(a)}\,.
\end{align}
Essentially, one chooses a tetrad basis such that the basis projects the relevant quantities on the original coordinate basis of $g_{\mu\nu}$ onto a particular basis of $\eta_{(a)(b)}$, which is commonly assumed to be the Minkowskian matrix
\begin{equation}
\eta_{(a)(b)}=\eta^{(a)(b)}=\textrm{diag}\left(-1,1,1,1\right)\,.
\end{equation}
Therefore, in the tetrad formalism, any vector or tensor field defined on the coordinate basis can be projected onto the tetrad frame, in which the fields can be expressed through their tetrad components:
\begin{align}
A_{\mu}&=e_{\mu}^{(a)}A_{(a)}\,,\quad A_{(a)}=e_{(a)}^{\mu}A_{\mu}\,,\nonumber\\
B_{\mu\nu}&=e_{\mu}^{(a)}e_{\nu}^{(b)}B_{(a)(b)}\,,\quad B_{(a)(b)}=e_{(a)}^{\mu}e_{(b)}^{\nu}B_{\mu\nu}\,.
\end{align}
On the other hand, the derivatives defined on the coordinate basis have to be manipulated with great care. Indeed, in the tetrad frame, the covariant (partial) derivative in the original coordinate frame is replaced with the intrinsic (directional) derivative. For instance, the derivatives of an arbitrary rank two object $H_{\mu\nu}$ in the two frames are related through the following equation:
\begin{align}
&\,H_{(a)(b)|(c)}\equiv e^{\lambda}_{(c)}H_{\mu\nu;\lambda}e_{(a)}^{\mu}e_{(b)}^{\nu}\nonumber\\
=&\,H_{(a)(b),(c)}\nonumber\\&-\eta^{(m)(n)}\left(\gamma_{(n)(a)(c)}H_{(m)(b)}+\gamma_{(n)(b)(c)}H_{(a)(m)}\right)\,,\label{2.7}
\end{align}
where a vertical rule and a comma denote the intrinsic and directional derivative with respect to the tetrad indices, respectively. A semicolon stands for a covariant derivative with respect to the tensor indices. On the above expression \eqref{2.7}, the Ricci rotation coefficients are defined by
\begin{equation}
\gamma_{(c)(a)(b)}\equiv e_{(b)}^{\mu}e_{(a)\nu;\mu}e_{(c)}^{\nu}\,.
\end{equation}
For more detail about the introduction and applications of the tetrad formalism, we refer the readers to Ref.~\cite{Chandrabook}.

At this point, we shall choose the following tetrad basis associated with the perturbed metric \eqref{perturbg}:{\footnote{The HMPG theory is a Lorentz invariant theory and the choice of the tetrad basis is not unique. However, the choice of Eq.~\eqref{tetradbasis111} is the most natural one and choosing other basis shall not change the result of the master equation.}}
\begin{align}
e^{\mu}_{(t)}&=\left(F^{-1/2},\quad\zeta F^{-1/2},\quad0,\quad0\right)\,,\nonumber\\
e^{\mu}_{(\varphi)}&=\left(0,\quad \frac{1}{r\sin\theta},\quad 0,\quad0\right)\,,\nonumber\\
e^{\mu}_{(r)}&=\left(0,\quad q_2G^{1/2},\quad G^{1/2},\quad0\right)\,,\nonumber\\
e^{\mu}_{(\theta)}&=\left(0,\quad q_3/r,\quad 0,\quad 1/r\right)\,,\label{tetradbasis111}
\end{align}
and
\begin{align}
e_{\mu}^{(t)}&=\left(F^{1/2},\quad0,\quad0,\quad0\right)\,,\nonumber\\
e_{\mu}^{(\varphi)}&=\left(-\zeta,\quad 1,\quad -q_2,\quad -q_3\right)r\sin\theta\,,\nonumber\\
e_{\mu}^{(r)}&=\left(0,\quad 0,\quad G^{-1/2},\quad0\right)\,,\nonumber\\
e_{\mu}^{(\theta)}&=\left(0,\quad 0,\quad 0,\quad r\right)\, ,\label{tetradbasis222}
\end{align}

\subsection{Master equation}
The field equation in HMPG, when written in the scalar-tensor representation, is given by Eqs.~\eqref{finalequationofmotionGT} and \eqref{eff}. In the tetrad frame, the field equation can be written as
\begin{align}
G_{(a)(b)}\left(1+\phi\right)=&-\frac{1}{2}\eta_{(a)(b)}\left(V+2\Box\phi\right)+\frac{3}{4\phi}\eta_{(a)(b)}\left(\partial\phi\right)^2\nonumber\\
-\frac{3}{2\phi}\phi_{,(a)}\phi_{,(b)}&+e^\mu_{(a)}\left(\phi_{,(b)}\right)_{,\mu}-\gamma_{(c)(b)(a)}\phi_{,(d)}\eta^{(c)(d)}\,.\label{tetradfieldeq}
\end{align}
Note that we have assumed a zero energy-momentum tensor. The master equation governing the gravitational perturbations is derived by linearizing the field equation \eqref{tetradfieldeq}. Using the tetrad basis given in Eqs.~\eqref{tetradbasis111} and \eqref{tetradbasis222}, the $(\theta,\varphi)$ and $(r,\varphi)$ components of the linearized field equation \eqref{tetradfieldeq} read
\begin{align}
R_{(\theta)(\varphi)}\left(1+\phi\right)&=-\gamma_{(r)(\varphi)(\theta)}\phi_{,(r)}\,,\nonumber\\
R_{(r)(\varphi)}&=0\,,\nonumber
\end{align}
which can be written explicitly as
\begin{align}
\left[\left(1+\phi\right)r^2\sqrt{FG}\sin^3\theta\left(q_{2,\theta}-q_{3,r}\right)\right]_{,r}&\nonumber\\
=\left(1+\phi\right)\frac{r^2\sin^3\theta}{\sqrt{FG}}&\left(\zeta_{,\theta}-q_{3,t}\right)_{,t}\,,\label{43.1}\\
\left[r^2\sqrt{FG}\sin^3\theta\left(q_{2,\theta}-q_{3,r}\right)\right]_{,\theta}&\nonumber\\
=r^4\sin^3\theta\sqrt{\frac{G}{F}}&\left(q_{2,t}-\zeta_{,r}\right)_{,t}\,,\label{43.2}
\end{align}
respectively. Defining the following new variable
\begin{equation}
\mathcal{Q}\equiv\left(1+\phi\right)r^2\sqrt{FG}\sin^3\theta\left(q_{2,\theta}-q_{3,r}\right)\,,
\end{equation}
and eliminating $\zeta$ in Eqs.~\eqref{43.1} and \eqref{43.2}, we get
\begin{align}
&\left(1+\phi\right)\left[\frac{\sqrt{FG}\mathcal{Q}_{,r}}{\left(1+\phi\right)r^2}\right]_{,r}\nonumber\\
&+\sqrt{\frac{F}{G}}\frac{\sin^3\theta}{r^4}\left(\frac{\mathcal{Q}_{,\theta}}{\sin^3\theta}\right)_{,\theta}=\frac{\mathcal{Q}_{,tt}}{r^2\sqrt{FG}}\,.\label{FGinterminda1}
\end{align}
Then, we consider the Fourier decomposition and the following ansatz:
\begin{equation}
\mathcal{Q}(r,\theta)=\mathcal{Q}(r)Y(\theta)\,,
\end{equation}
where $Y(\theta)$ is the Gegenbauer function  \cite{Abramow} and it satisfies
\begin{equation}
\frac{d}{d\theta}\left(\frac{1}{\sin^3\theta}\frac{dY}{d\theta}\right)=-\left[l(l+1)-2\right]\frac{Y}{\sin^3\theta}\,,
\end{equation}
with $l$ being the multipole number. Using the above definitions, Eq.~\eqref{FGinterminda1} can be rewritten as
\begin{align}
&\left(1+\phi\right)\left[\frac{\sqrt{FG}\mathcal{Q}_{,r}}{\left(1+\phi\right)r^2}\right]_{,r}\nonumber\\
&-\sqrt{\frac{F}{G}}\frac{\left[l(l+1)-2\right]}{r^4}\mathcal{Q}=-\frac{\omega^2\mathcal{Q}}{r^2\sqrt{FG}}\,.\label{FGinterminda2}
\end{align}

Finally, we define
\begin{equation}
\psi_g\equiv\mathcal{Q}/\mathcal{X}\,,
\end{equation} 
where $\mathcal{X}\equiv r\sqrt{1+\phi}$, and use the tortoise radius $r_*$ defined as
\begin{equation}
\frac{dr}{dr_*}=\sqrt{FG}\,.
\end{equation}
The master equation \eqref{FGinterminda2} can be written in a Schr\"odinger-like form:
\begin{equation}
\partial_{r_*}^2\psi_{g}+\omega^2\psi_g=V_g(r)\psi_g\,,\label{masterequationgg}
\end{equation}
where the effective potential $V_g(r)$ reads{\footnote{Although in this paper, the master equation \eqref{masterequationgg} is derived within the scalar-tensor representation, it can also be derived directly by using Eqs.~\eqref{fieldequation}, \eqref{fieldequation2}, and \eqref{compatible2}.}}
\begin{equation}
V_g(r)=F\left[\frac{l(l+1)-2}{r^2}-\mathcal{X}\sqrt{\frac{G}{F}}\left(\frac{\sqrt{FG}\mathcal{X}_{,r}}{\mathcal{X}^2}\right)_{,r}\right]\,.\label{Vg}
\end{equation}

The master equation \eqref{masterequationgg} and the potential \eqref{Vg} are the main results of this paper. The master equation governs the axial perturbations of a static and spherically symmetric spacetime in vacuum HMPG. Note that in the derivation of the master equation, we do not make the assumption that the potential $V(\phi)$ must be zero. In fact, it can be seen that the expression of the master equation does not contain explicitly the scalar field potential. However, the inclusion of the scalar field potential would affect the QNMs implicitly. More precisely, changing the scalar field potential would affect the behaviors of the background spacetime, that is, the metric functions $F$ and $G$, as well as the scalar field $\phi$. Therefore, the QNMs would be altered as well.

Moreover, if we assume that the scalar field is simply a constant, namely, the dynamics of the scalar field is turned off, the metric functions $F$ and $G$ would reduce to those of the Schwarzschild-(A)dS metric, in which the effective cosmological constant is determined by the value of the scalar field potential. Also, the effective potential \eqref{Vg} would reduce to its GR counterpart and the master equation reduces to the well-known Regge-Wheeler equation.

\begin{figure}[!ht]
\centering
\vspace{-3pt}
\includegraphics[width = .45\textwidth]{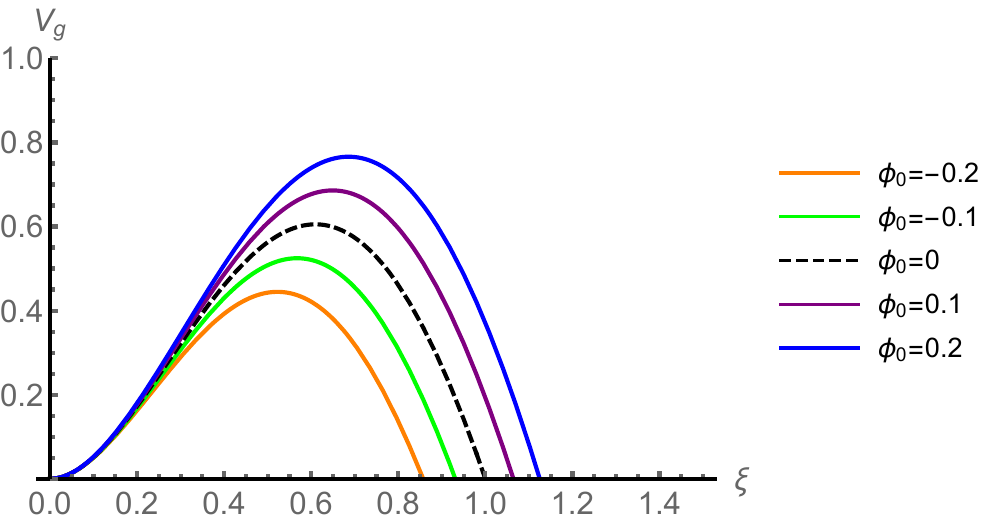}
\vspace{-10pt}
\caption{\label{potentialVG}The effective potential $V_g$ given by Eq.~\eqref{Vg} as a function of $\xi$. The black dashed curve shows the Regge-Wheeler potential for the Schwarzschild black hole. The colored curves show how the values of $\phi_0$ alter the shape and height of the effective potential.}
\end{figure}

In FIG.~\ref{potentialVG}, we assume some values of $\phi_0$ and present the effective potential \eqref{Vg} as a function of $\xi=1/r$. The values of $u$ at the asymptotic region is approximated by $\phi_0$ times a very small constant, say, $10^{-11}$, as we have mentioned at the end of Sec.~\ref{sec.BH}. One can see that the potential vanishes at two points, one is at the spatial infinity ($\xi\rightarrow0$) and the other is at the event horizon where $\xi$ can vary around unity, depending on the values of $\phi_0$. Essentially, having a positive (negative) $\phi_0$ would effectively suppress (enhance) the gravitational field around a local gravitating object (see the effective gravitational constant given by Eq.~\eqref{Geffconstant} and the effective energy-momentum tensor \eqref{eff}). In this sense, the radius of the event horizon would be smaller (larger) than the Schwarzschild counterpart. This is consistent with the results in Ref.~\cite{Danila:2018xya}. For calculating the QNM frequencies, we only need to consider the exterior spacetime, that is, the spacetime region from the event horizon up to spatial infinity. This corresponds to the domain of $\xi$ where $V_g \ge 0$ and we have only focused on this region of $\xi$ in FIG.~\ref{potentialVG}. Note that we have rescaled all the quantities by assuming $2M=1$. The black dashed curve in FIG.~\ref{potentialVG} represents the Regge-Wheeler potential for the Schwarzschild black hole, in which $\phi_0=u=0$.

\subsection{QNM frequencies}
The master equation \eqref{masterequationgg} with the potential given by Eq.~\eqref{Vg} describes the axial perturbations of a static and spherically symmetric black hole in HMPG. In order to compute the QNM frequencies, the master equation should be treated as an eigenvalue problem with appropriate boundary conditions. For an asymptotically flat and isolated black hole, the boundary conditions subject to the system are that only outgoing waves appear at spatial infinity and there are only ingoing waves moving toward the black hole at the event horizon. Generally speaking, the QNM frequencies of a black hole are discrete and complex-valued. In this paper, we will use a semianalytical approach, which is based on the WKB approximation, to calculate the QNM frequencies.

The WKB method for calculating QNM frequencies was first formulated in the seminal paper \cite{Schutz:1985km} and was then improved to higher orders in Refs.~\cite{Iyer:1986np,Konoplya:2003ii,Matyjasek:2017psv,Matyjasek:2019eeu,Hatsuda:2019eoj} (see Ref.~\cite{Konoplya:2019hlu} for the recent review about the WKB method). The advantage of the WKB method is that the QNM frequencies can be directly evaluated using a simple formula, as long as the effective potential in the master equation is provided. Although the WKB method is merely a semianalytic approach, it is proven to be accurate when the multipole number $l$ is larger than the overtone $n$ \cite{Berti:2009kk}. This happens to be propitious from the astrophysical point of view in the sense that the fundamental QNMs with $n=0$ typically decay more slowly, hence they are more detectable than other overtones. Moreover, for a merger event of a binary black hole system, the modes with $l=2$ or $l=3$ have larger magnitudes, compared with other high-$l$ modes. Therefore, in this paper we will focus on the fundamental modes with $l=2$ and $l=3$. 

We use the 6th order WKB method in which the QNM frequencies can be obtained from the following formula \cite{Konoplya:2003ii}
\begin{equation}
\frac{i\left(\omega^2-V_{gm}\right)}{\sqrt{-2V^{(2)}_{gm}}}-\sum^6_{i=2}\Lambda_i=n+\frac{1}{2}\,,
\end{equation}
where $V_{gm}$ is the maximum value of the potential $V_g$, and $V^{(2)}_{gm}$ is the second derivative of $V_g$ with respect to $r_*$ evaluated at the potential maximum. The constant coefficients $\Lambda_i$ are related to higher order WKB corrections and they are given in Refs.~\cite{Iyer:1986np,Konoplya:2003ii}.

\begin{figure}[h]
\centering
\vspace{5pt}
\includegraphics[width = .45\textwidth]{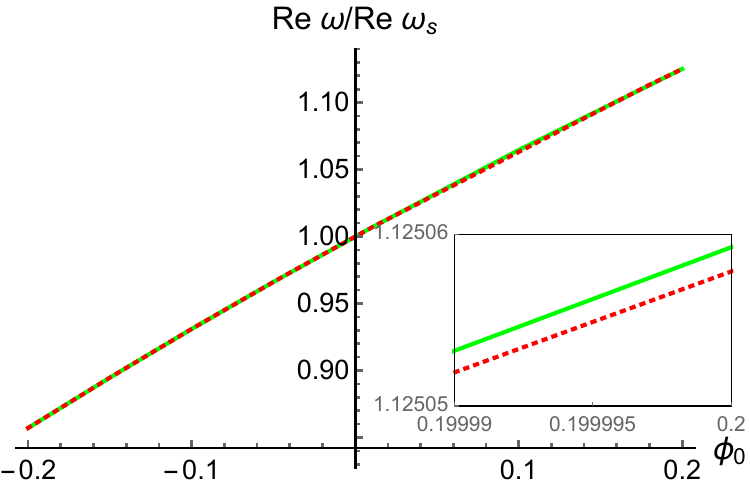}
\includegraphics[width = .45\textwidth]{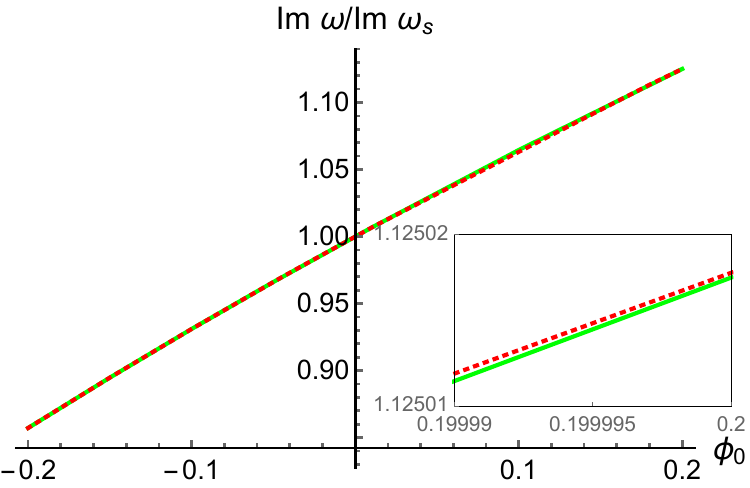}
\vspace{5pt}
\caption{\label{qnmresult}The real part (upper) and the imaginary part (lower) of the fundamental QNM frequencies are presented with respect to $\phi_0$. The green and the dotted red curves represent the results of $l=2$ and $l=3$, respectively. Note that in this figure, we consider the frequency ratio associated with the Schwarzschild black hole. The subfigures show the detailed behavior of the curves near $\phi_0\approx0.2$, where the difference between the red and green lines is expected to be maximized. The relative difference between the two lines is of the order of $10^{-6}$ $(10^{-7})$ for the real (imaginary) part.}
\end{figure}

In the upper panel of FIG.~\ref{qnmresult}, we consider the fundamental modes and show the ratio between the real part of the QNM frequencies for the HMPG black holes and that of the Schwarzschild black hole $\omega_s$. The imaginary part of the QNMs are shown in the lower panel. In both panels, the green and the dotted red curves represent the results of $l=2$ and $l=3$, respectively. It can be seen that the results (the changes with respect to $\omega_s$) corresponding to these two multipole numbers are nearly identical in the chosen range of the asymptotic scalar field $\phi_0$, which is required to be small subject to the post-Newtonian constraints.

According to FIG.~\ref{qnmresult}, one can see that both the real part and the absolute value of the imaginary part of the QNM frequencies would increase when $\phi_0$ increases. Also, when $\phi_0=0$, the frequency reduces to that of the Schwarzschild black hole. This is expected because we have assumed that the asymptotic value of $u$ is proportional to $\phi_0$ (times a very tiny constant). When $\phi_0=0$, the asymptotic value of $u$ vanishes and the dynamics of the scalar field is turned off. Note also that the potential $V_g$ in the master equation reduces to the standard Regge-Wheeler potential when $\phi_0=0$ (see the black dashed curve in FIG.~\ref{potentialVG}).

\section{Conclusions}\label{sec.conclu}
The HMPG theory can be regarded as a combination of the metric-$f(R)$ gravity and the Palatini-$f(R)$ gravity. When written in its scalar-tensor representation, it can be shown that the scalar field is dynamical and the theory is able to describe the late-time acceleration of the universe. In addition, the theory can pass the local experimental tests at the Solar System scale without invoking any screening mechanism. Such mechanism is commonly required in many modified theories of gravity with infrared corrections, such as the metric-$f(R)$ gravity. Also, unlike the Palatini-$f(R)$ gravity, the HMPG theory does not suffer from the microscopic instabilities, which appear due to unwanted non-trivial matter-curvature couplings. 

In the presence of the dynamical scalar field, the vacuum black hole solutions in the HMPG theory could differ from their GR counterparts and provide us with an opportunity to distinguish them by investigating their QNM spectra. In this paper, we focus on the axial gravitational perturbations and derive the master equation governing the perturbations. The master equation is obtained by using the tetrad formalism and it can be written as a single differential equation in a Schr\"odinger-like form, indicating that the axial gravitational perturbations do not couple with the additional degree of freedom. However, the dynamical scalar field which appears at the unperturbed level would affect the master equation and change the QNM spectra. According to the post-Newtonian constraints, the value of the scalar field at the asymptotic region, namely, $\phi_0$, is required to be very small. In the investigation of the black holes and their QNMs, we take these constraints into account and see how the QNM frequencies deviate from those of the Schwarzschild black hole in the observationally consistent range of parameter space. We find that, in the range of the parameter space under consideration, increasing the value of $\phi_0$ would increase the real part and the absolute value of the imaginary part of the QNM frequencies. 

Typically, the gravitational perturbations of a black hole consist of the axial modes and the polar modes. For the Schwarzschild black hole in GR, the isospectrality between these two modes, i.e., the polar and the axial modes share the identical spectrum, is a very unique feature. Any evidence of the isospectrality breaking for a Schwarzschild black hole would be a smoking-gun of going beyond GR. For the HMPG theory, as can be seen from Eq.~\eqref{masterequationgg}, the master equation for the axial modes is source-free and it does not couple to the scalar mode corresponding to the dynamical scalar field. Furthermore, when the dynamics of the scalar field $\phi$ at the background level is turned off, the Regge-Wheeler equation is recovered. As for the polar modes, although a thorough scrutiny is beyond the scope of the present paper and is going to be carried out elsewhere, it can be expected that the additional scalar mode would couple to the polar modes by sourcing the master equation of the latter. The source term is expected to be present even though the scalar field at the background level is non-dynamical. In particular, when one considers the gravitational perturbations of a Schwarzschild black hole in HMPG, the isospectrality between the polar modes and the axial modes would be broken because the additional source term in the polar sector would drive new perturbation modes \cite{Qian:2020wbv}. Another theoretical example for the breaking of isospectrality is the metric-$f(R)$ gravity \cite{Bhattacharyya:2017tyc,Bhattacharyya:2018qbe,Datta:2019npq}. Essentially, by examining the way how the isospectrality is broken in HMPG and comparing it with those in other gravitational theories, one can use it as a promising tool, in addition to the direct comparison of the QNM spectra, to test various theoretical models. We leave these issues to our future investigations.

\acknowledgments

CYC would like to thank Tiberiu Harko for fruitful comments and instruction. CYC, YHK, and PC are supported by Ministry of Science and Technology (MOST), Taiwan, through No. 107-2119-M-002-005, Leung Center for Cosmology and Particle Astrophysics (LeCosPA) of National Taiwan University, and Taiwan National Center for Theoretical Sciences (NCTS). CYC is also supported by MOST, Taiwan, through No. 108-2811-M-002-682 and Institute of Physics of Academia Sinica. PC is in addition supported by US Department of Energy under Contract No. DE-AC03-76SF00515.

\end{document}